\documentclass[12pt]{article}
\usepackage{a4}
\usepackage{amsfonts}
\usepackage{amssymb}
\usepackage{cite}
\usepackage{epsfig}
\usepackage{amsmath}

\usepackage[english]{babel}

\author{}

\newcommand{\drawsquare}[2]{\hbox{%
\rule{#2pt}{#1pt}\hskip-#2pt
\rule{#1pt}{#2pt}\hskip-#1pt
\rule[#1pt]{#1pt}{#2pt}}\rule[#1pt]{#2pt}{#2pt}\hskip-#2pt
\rule{#2pt}{#1pt}}
\newcommand{\fund}{\raisebox{-.5pt}{\drawsquare{6.5}{0.4}}}
\newcommand{\Ysymm}{\raisebox{-.5pt}{\drawsquare{6.5}{0.4}}\hskip-0.4pt%
         \raisebox{-.5pt}{\drawsquare{6.5}{0.4}}}
\newcommand{\Yasymm}{\raisebox{-3.5pt}{\drawsquare{6.5}{0.4}}\hskip-6.9pt%
        \raisebox{3pt}{\drawsquare{6.5}{0.4}}}
\newcommand{\antifund}{\overline{\fund}}

\newcommand{\beq}{\begin{equation}}
\newcommand{\eeq}{\end{equation}}
\newcommand{\ba}{\begin{array}}
\newcommand{\ea}{\end{array}}
\newcommand{\bea}{\begin{eqnarray}}
\newcommand{\eea}{\end{eqnarray}}

\newcommand{\ov}{\overline}
\def\IR{\relax{\rm I\kern-.18em R}}

\def\IP{\relax{\rm I\kern-.18em P}}
\def\inbar{\vrule height1.5ex width.4pt depth0pt}
\def\IC{\relax\,\hbox{$\inbar\kern-.3em{\rm C}$}}

\def\K3{{\bf K3}}

\def\th{\theta}

\def\ov{\overline}

\def\n2d{\cN_{V^*}^{\otimes 2}}

\def\IT{\bf T}

\def\IC{\mathbb{C}}

\def\IR{\mathbb{R}}

\def\IZ{\mathbb{Z}}

\def\IP{\mathbb{P}}

\def\cN{{\mathcal N}}

\def\to{\rightarrow}

\def\beq{\begin{equation}}
\def\eeq{\end{equation}}
\def\beqa{\begin{eqnarray}}
\def\eeqa{\end{eqnarray}}

\begin{document}

\title{
On the (Non) Perturbative Origin of Quark \\ Masses in D-brane GUT Models}
 \vspace{1.0cm}
\author{  Christos Kokorelis$^1$}

\date{}

\maketitle

\begin{center}
\emph{${}^1$ Institute of Nuclear Physics, N.C.S.R. Demokritos, 15310, Athens, Greece}
\vspace{0.2cm}

\tt{kokorelis@inp.demokritos.gr }
\vspace{1.0cm}
\end{center}
\vspace{0.7cm}

\begin{abstract}

We examine the issue of generating the perturbatively absent
${\bf 10} \cdot {\bf 10} \cdot { {\bf 5}}^H$ SU(5)/flipped SU(5)Yukawa couplings
in type II D-brane orientifold compactifications of string theory
both at the perturbative (PER) and the non-perturbative (NP) level.
We find at the PER level, higher order terms like
$ {\bf 10} \cdot {\bf 10} \cdot
{\bf{\bar 5}}^H \cdot{\bf{\bar 5}}^H \cdot {\bf{\bar 5}}^H \cdot {\bf{\bar
    5}}^H \cdot {\bf 1}^H \cdot {\bf 1}^H$ in SU(5) may be responsible for the relevant quark mass generation in models with general intersecting D6-branes.
Euclidean D2-brane instantons on the other hand
can also generate at the NP via the
term ${\bf 10} \cdot {\bf 10} \cdot
{\bf{\bar 5}}^H \cdot{\bf{\bar 5}}^H \cdot {\bf{\bar 5}}^H \cdot {\bf{\bar 5}}^H $
the relevant quark masses by the use of just the $U(1)_b$ brane, for
SU(5) and flipped SU(5) GUTS classes of models. We provide local examples of rigid $O(1)$ instantons within the $T^6/{\mathbb Z}_2 \times {\mathbb
Z}_2'$ toroidal orientifold with torsion, whose NP contribution to the
masses gets minimal as it is induced by just a duplicated
disk diagram.

\end{abstract}

\thispagestyle{empty} \clearpage

\tableofcontents

\section{Introduction}

Intersecting brane worlds (see \cite{ur1,kokos1,kir1,lust1,blum1,blum2} for reviews) provide us with a beautiful arena that has become in recent years the main playground for realistic model building (RMB) attempts. The main characteristics of these models against previous attempts in the context of RMB in string theory is the appearance of the right handed neutrino in the massless spectrum localized at the string scale. Often the baryon mumber B or B-L are gauged symmetries, hence they survive as global symmetries to low energies.
This happens since the $B \wedge F_a$ couplings, that take part in a generalized Green-Schwarz mechanism, give masses to the associated U(1) gauge bosons $F_a$.
Moreover, in orientifold compactifications of IIA (or IIB) in four dimensions, one finds that intersecting D-brane models possess
perturbatively absent matter couplings (PAMC) \cite{hep-th/0107138} that
has singled out all the SU(5)-like GUTS. In this case,
the relevant Yukawa terms
\begin{align}
\langle {\bf 10}_{(2,0)} \cdot {\bf 10}_{(2,0)} \cdot  {\bf 5}^H_{(1,1)} \rangle,
\label{as1}
\end{align}
[see also section 2] are not allowed in SU(5)-like GUTS from intersecting branes due to charge non-conservation. Solutions to PAMC could be provided by the existence of instantons generating the relevant couplings.
Because stringy instantons
break global U(1) symmetries they can generate PAMC. Hence they have been used recently to generate various types of non-perturbative effects
\cite{ins1a,Witten:19,Dine:1986zy,Wit,Ganor:1996pe,ins1,ins2,ins3,Haack:2006cy,Abel:2006yk,Akerblom:2006hx,Bianchi:2007fx,Cvetic:2007ku,Argurio:2007qk,Argurio:2007vq,Bianchi:2007wy,isu,Akerblom:2007uc, Antusch:2007jd,Blumenhagen:2007zk,Aharony:2007pr,Aharony:2007db,Blumenhagen:2007bn,Billo:2007sw,Aganagic:2007py,Camara:2007dy,Cvetic:2007qj,Ibanez:2007tu,GarciaEtxebarria:2007zv,
Petersson:2007sc,Blumenhagen:2007sm,Bianchi:2007rb, Blumenhagen:2008ji,Matsuo:2008nu,Argurio:2008jm,Cvetic:2008ws,Cvetic:2008hi,GarciaEtxebarria:2008pi,Buican:2008qe,Forcella:2008au,Blumenhagen:2008kq,Camara:2008zk,Billo':2008sp,Billo':2008pg,
Kumar:2008cm,Marsano:2008jq,Marsano:2008py,Uranga:2008nh,Heckman:2008qt,Akerblom:2007nh, Cvetic:2007sj,Billo:2002hm,Ibanez:2008my}.
Instanton induced phenomenological couplings in the literature exist for the Majorana \cite{ins1,ins2,ins3}/ Dirac \cite{Cvetic:2008hi} masses for neutrinos,
$\mu$-terms \cite{ins2}, an instanton proposal \cite{Blumenhagen:2007zk} for the long standing problem of the perturbatively absent Yukawa coupling (\ref{as1})
of SU(5)-like GUTS \cite{hep-th/0107138} etc.
In IIA compactifications with D-branes, the relevant non-perturbative effects are induced from Euclidean D2-brane instantons. The E2-instanton, wraps a three-cycle $\Pi_{E_2}$ and under a $U(1)_a$ transformation the instanton action $S_{E_2}$ transforms as

\beqa
e^{-S_{E_2}} &=& e^{\left[\frac{2\pi}{l_s^3} ( -\frac{1}{g_s}Vol_{\Pi_{E_2}} + i \int_{\Pi_{E_2}} C^{(3)} ) \right]}
\rightarrow \Lambda_{\Pi_{E_2}},\nonumber\\
\Lambda_{\Pi_{E_2}}&=& e^{Q_a(E_2)} e^{-S_{E_2}},
\label{insta1}
\eeqa
\begin{align}
Q_a(E_2) = -N_a \ \Pi_{E_2} * [\Pi_a -\Pi_a^{\prime}]
\label{insta2}
\end{align}
where the D6-brane wraps the three-cycle $\Pi_a$ and its orientifold image $\Pi_a^{\prime}$.
Chiral fermions appears as open strings stretching between the two intersecting branes, while the number of chiral fermions between an instanton and a D6-brane $\alpha$ is described by the intersection number $I_{E2 a}$.
"Charged" fermionic zero modes appear at the intersection of E2 and a D6-brane $\alpha$. If $ \Pi_i \Phi_i$ is the coupling, made of a product of $i$ chiral superfields, that is not invariant under a global symmetry then the instantons induce F-term couplings such that the superpotential term $W = \Pi_i \Phi_i
e^{-S_{E_2}^{cl}}$ is invariant under the global symmetries.
For {\cal O}(1) instantons wrapping a rigid, orientifold invariant cycle in the internal manifold the charge carried by
$Q_a(E_2) = -N_{\alpha} \Pi_{E_2} * [\Pi_{\alpha}]$  is exactly the amount of U(1)$_{\alpha}$ zero modes carried by the fermionic zero modes between $E_2$ and the D6$_{\alpha}$. To establish notation a positive intersection number associates to $I_{E2 a}$ the transformation representation behaviour $({\fund_{E_2}, \antifund}_a)$.

In this work, we investigate the generation of the SU(5) up-quark mass coupling (\ref{as1})), that was known to be absent perturbatively in all D-brane SU(5) GUTS from intersecting brane worlds (IBW's), either in IIA(or in IIB), by finding a new type of SU(5) gauge group higher, than trilinear, invariant perturbative Yukawa coupling in the form
\begin{align}
\frac{1}{M_s^{N+3}}  {\bf 10} \cdot {\bf 10} \cdot {\bf \overline{5} }^H \cdot {\bf \overline{5}}^H \cdot {\bf \overline{5} }^H  \cdot {\bf \overline{5} }^H \cdot \Phi_1^H  \cdots \Phi_N^H \ ,
\label{newco}
\end{align}
where $\Phi_i$ gauge singlet Higgs fields.
 In section 2 we present the general form of the
higher dimensional perturbative coupling responsible for generating
the up-quark mass based on eqn.(\ref{newco}). We exhibit the presence of this term
in a non-supersymmetric background.
Thus all SU(5) models from intersecting branes receive, beyond tree level, non-zero corrections to the mass of the up-quarks.
 In section 3, we exhibit the presence of this perturbative coupling in a N=1 supersymmetric $Z_2 \times Z_2$ orientifold. In section 4, we discuss a new form of  instanton coupling which contributes to the mass of the
 up-quarks and originates from the perturbative coupling of sections 2 and 3.
We also present a local realization based on the $T6/(\IZ_2 \times \IZ_2^{\prime})$ orientifold with discrete
torsion \cite{Blumenhagen:2005tn}, in which the instanton contribution to the coupling (\ref{as1}) gets generated by a rigid O(1) instanton.

\section{Perturbative induced quark masses for SU(5) D-brane
models}

\subsection{The perturbative term in D-brane models}
Model building based on SU(5)(or flipped SU(5)) GUT models from 4D type IIA compactifications, has been hampered as in these models the missing up(down) quark masses excluded the models phenomenologically. The first semirealistic SU(5) models historically were constructed in \cite{Bachas:1995ik} (see also some recent work on \cite{Bachas:2008jv}, even though in a language T-dual to the one used in intersecting brane worlds(IBW)). In the context of IBW a two stack three generation non-susy SU(5) based on Z3 orientifolds appeared in \cite{hep-th/0107138}, where the existing Yukawa couplings were listed.
Also a toy four stack N=1 SU(5) model has been constructed on Z2 x Z2 orientifolds in \cite{hep-th/0107166}.
Moreover, in \cite{eknano} it was noticed that by a simple rescaling of the massless U(1) surviving the Green-Schwarz mechanism, the models of \cite{hep-th/0107138} were converted into flipped SU(5) ones, while it also appeared that the models were
missing also the GUT Higgses. In \cite{afk} we proposed in which way one can identify GUT Higgses in a general non-supersymmetric flipped SU(5) model coming from intersecting branes and also we
identified all the proton decay modes and presented a new doublet-triplet splitting mechanism.
 SU(5) models based on intersecting branes have been also constructed
in different contexts (e.g. in the presence of fluxes; see also \cite{Kokorelis:2004gb}) or in the context of other orientifold compactifications however, the basic spectrum structure is the one identified in \cite{hep-th/0107138}, \cite{eknano}, \cite{afk}.

We note that the matter field content (MFC) of an SU(5) GUT is
\beq
{\bf {\ov{10}}} = (Q, u^c, e^c), \ {\bf {\bar 5}} = (d^c, L),  \ {\bf 1} = \nu^c ,
\eeq
while the MFC of a flipped SU(5) GUT reads
\beq
{\bf {\ov{10}}}_1 = (u, d, d^c, \nu^c), \ {\bf {\bar 5}}_{-3} = (d^c, L),  \ {\bf 1}_5 = e^c .
\eeq
In the minimal string version \cite{hep-th/0107138}, \cite{eknano}, \cite{afk}, of the SU(5) GUT models there are two stacks a, b, of intersecting D6-branes,
associated to U(5)$_a$, U(1)$_b$ branes respectively. Its chiral spectrum can be seen in table (\ref{tablegut}). We note that the chiral spectrum of table (\ref{tablegut}) can be embedded also in a general string construction and thus our discussion in this section is quite generic. When embedded in a different string construction, the D6-branes giving rise to the
chiral spectrum of table (\ref{tablegut}) may be accompanied by a number of extra space filling D6-branes necessary to cancel tadpoles (RR tadpoles in a non-susy model; RR and NSNS tadpoles
for N=1 models).
To establish notation, we note that
the multiplet ${\bf 10}_{(2,0)}$ also contains the GUT Higgs field which should appear as a vector-like pair as it has been firstly noted in \cite{afk}.
The massless U(1) surviving massless the Green-Schwarz mechanism is given by U(1)$_X$ seen in table (\ref{tablegut}) ($U(1)_Y$ in flipped SU(5)). In the ab$^{\prime}$ sector where $I_{ab^{\prime}}
 = 0 $ there are present the non-chiral Higgs $5, {\bar 5}$. If the SU(5) model is non-supersymmetric then the Higgs fields are coming from the lowest massive excitation spectrum of the fermions $5, {\bar 5}$ with the same U(1)
 charges (see \cite{afk} for details and also table (\ref{tablegut})).

\begin{table}[ht]
\centering
\begin{tabular}{|c|c|c|c|c|}
\hline
sector & number &  $U(5)_a\times U(1)_b$ reps. & $U(1)_X$  & $U(1)_Y$ \\
\hline \hline
$(a,a')$ &  $3$ & ${\bf 10}_{(2,0)}$   & $\frac{2}{5}$  & $1$\\
$(a,b)$ &  $3$ & $\ov{\bf 5}_{(-1,1)}$   & $-\frac{6}{5}$ & $-3$ \\
$(b,b')$ &  $3$ & ${\bf 1}_{(0,-2)}$ & ${2}$ &  $5$   \\
$(a,a')$ &  $1$ & ${\bf 10}^H_{(2,0)} +\ov{\bf 10}^H_{(-2, 0)}  $ &  $\frac{1}{2}$   &  $(1)  +  (-1)$\\
$(a,b')$ &  $1$ & ${\bf 5}^H_{(1,1)}+\ov{\bf 5}^H_{(-1,-1)}$ & $(-1) + (1)$  & $(-2)  +  (2)$ \\
\hline
\end{tabular}
\caption{$SU(5)$ GUT via intersecting D6-branes : The massless U(1), $U(1)_X = \frac{1}{5} U(1)_a - U(1)_b$ (in a flipped SU(5) parametrization $U(1)_Y = \frac{1}{2} U(1)_a - \frac{5}{2}U(1)_b$). The first three rows describe the chiral fermionic spectrum. The last two rows
contains the matter Higgses: electroweak 5's, ${\bar 5}$'s and GUT 10, $\ov{10}$'s .
\label{tablegut} } 
\end{table}

The Yukawa couplings giving masses to the down-quarks(up-quarks) for SU(5) (flipped SU(5)) respectively are given by the tree level expression ${\bf 10} \cdot {\bf{ \bar 5}} \cdot {\bf {\bar 5}}^H$.
On the contrary the masses for up (down) quarks in SU(5)( flipped SU(5)),
are given by the coupling (\ref{as1}),
 which is not allowed by charge conservation in orientifold compactifications of string theory. It violates the charge conservation by the amount
 (U(1)$_a$, U(1)$_b$) = (5, 1) units.
In this work, our first result is that the coupling (\ref{as1}) can be generated perturbatively.
One can imagine that the simplest solution to the missing mass term (\ref{as1})
may be coming from the potential term
\beqa
\frac{1}{M_s^7} \langle {\bf 10}_{(2,0)} \cdot {\bf 10}_{(2,0)} \cdot {\bf 5}^H_{(1, 1)}
\cdot {\bf {\bar 5}}^H_{(-1,-1)} \cdot
{\bf {\bar 5}}^H_{(-1,-1)}
 \nonumber\\
  \cdot {\bf {\bar 5}}^H_{(-1,-1)}   \cdot {\bf {\bar 5}}^H_{(-1,-1)}
\cdot {\bf {\bar 5}}^H_{(-1,-1)}
\cdot {\bf {\bar 1}}_{(0,2)}^H \cdot {\bf {\bar 1}}_{(0,2)}^H \rangle . && \nonumber\\
\label{koa2}
\eeqa
However a closer look reveals that a lowest order term exists, able to generate the relevant quark masses, via the Yukawa coupling
\beqa
\frac{1}{M_s^5} \langle {\bf 10}_{(2,0)} \cdot {\bf 10}_{(2,0)} \cdot {\bf {\bar 5}}^H_{(-1,-1)} \cdot {\bf {\bar 5}}^H_{(-1,-1)} \cdot {\bf {\bar 5}}^H_{(-1,-1)} \nonumber\\
\cdot {\bf {\bar 5}}^H_{(-1,-1)}
  \cdot {\bf {\bar 1}}_{(0,2)}^H \cdot {\bf {\bar 1}}_{(0,2)}^H \rangle .
\label{koa1}
\eeqa
The above term is gauge and charge invariant. Its gauge invariance is easily seen via the SU(5) tensor products
\beqa
{\bf 10} \cdot {\bf 10} \cdot {\bf {\bar 5}} \cdot {\bf {\bar 5}} \cdot {\bf {\bar 5}}
\cdot {\bf {\bar 5}} \sim {\bf {\bar 5}} \cdot {\bf {\bar 5}} \cdot {\bf {\bar 5}}
\cdot {\bf {\bar 5}} \cdot {\bf {\bar 5}} \sim {\bf {\ov {10}}} \cdot
{\bf {\bar 5}} \cdot {\bf {\bar 5}} \cdot {\bf {\bar 5}} \nonumber\\
\sim {\bf 10} \cdot \cdot {\bf {\bar 5}} \cdot {\bf {\bar 5}}
\eeqa
The term of eqn.(\ref{koa1}) cannot be used in the "re-scaled" two
stack version of the present model that converts it to its flipped SU(5) version as the singlet in this case accommodates the right handed electron and cannot get a vev.
Instead one might try to built a higher than two stack SU(5) GUT that has extra singlets to generate the required invariant in (\ref{koa1}).
The bosons associated to the ${\bf {1}}_{(0,2)}$ for the non-susy
model are part of the massive spectrum that is localized in the bb$^{\prime}$ intersection, that allows as a Higgs field the vector-like pair ${\bf 1}_{(0,2)}^H \ + \ {{\bf {\bar 1}}}_{(0, -2)}^H$ (see \cite{afk} and the 2nd ref. of \cite{kokos1} for details).
After the associated Higgses receive vevs, the order
of the quark masses is given by the perturbative evaluation of the above correlator
which reads
\begin{align}
m_U = Y_U \frac{1}{M_s^5} \ {\langle {\bf 5}^H \rangle}^4 \ {\langle {\bf 1}^H \rangle}^2 e^{- {\frac{1}{2 \pi \alpha^{\prime}}} A} ,
\label{exp}
\end{align}
where A is the worldsheet area of the 8-point correlator in
 (\ref{koa1})(see also \cite{Abel:2003vv}).

By looking at the formulae (\ref{exp}) we see that the derived U-quark masses depend on three quantities, the string scale $M_s$, the vev of 5-plet Higgses, and the area suppression factor A. Before we give an estimate of the U-mass, let us make some remarks about these quantities.

{\em With a High string scale}:
\newline
The string scale in the present models and in general in models where orientifolded intersecting D6-branes wrap a six dimensional toroidal internal space \cite{kokos1}, \cite{Bachas:1995ik}, \cite{hep-th/0107138}, \cite{afk}, \cite{hep-th/0107166}
 takes high values. On the other hand, the vev of the 5-plet Higgses can be taken to be (see N=1 examples in the next section) either at the electroweak scale $\upsilon = 246$ GeV or high at its natural value that could be the string scale $M_s$. In the previous case, namely Case I in eqn. (\ref{esti}),  the mass of the U-quark is very suppressed and is excluded phenomenologically as a typical suppression factor
can be of the order of $10^{-43} \times Exp[-M_s^2 A]$ GeV with $M_s \approx 10^{16}$ GeV and $\langle {\bf \overline{5}} \rangle = \upsilon$. In the latter case, case II, the natural vev of the 5-plet higgses is
of order $M_s$ and hence the smallness of the up(down) quark masses
is achieved from the exponential area suppression factor. The relevant estimates are summarized in eqn. (\ref{esti}). We assume
that the areas of the second and third tori are close to zero and the eight term coupling can
be approximated as
$M_U \sim  e^{- R_1 R_2 M_s^2 A^{(8)} } \equiv e^{-{\tilde A}}$
where the A-area may be of order one in string units.

\begin{align}
  m_U  \stackrel{High \ M_s}{\sim}  \left\{  \begin{array}{lll}
         M_w^4 \  M_s^{-3}\ e^{-{\tilde A}} & \mbox{if $\langle {\bf 5 }^H \rangle \sim \upsilon$,   \ $\langle {\bf 1 }\rangle  \sim M_s $: \ }&  \mbox{ Case I}
         ;\\\nonumber\\ & & \\\nonumber\\
        M_s \ e^{-{\tilde A}} & \mbox{if $\langle {\bf 5 }^H \rangle \sim \langle {\bf 1 }^H \rangle \sim M_s $:  \ }&\mbox{ Case II },  \\\end{array}
          \right .
          \end{align}
\beq
\label{esti}
\eeq
The value of the up-quark as a current quark mass estimate (see p.479 of \cite{rev}) in the $\overline{MS}$,
is between 1.5 - 4.0 MeV.
Hence, in Case II for a typical value of the string scale $M_s = 10^{16}$ GeV,
the mass of the up-quark e.g. 2 MeV is achieved for a suppression area factor "area" ${\tilde A} \approx 18.7$.

Summarizing, case I is excluded in models with a high string scale and the assumptions made in eq. (\ref{esti}). On the contrary,
case II as it appears in eq.(\ref{esti}) is valid in D-brane models with a high and also a low string scale as it is shown next.
In the following, we numerically examine under which conditions cases I, II are also valid in models with two extra dimensions, where we re-named them as Cases III, IV, respectively.

{\em With a Low string scale; Models from Extra Dimensions}:\newline
In type I compactifications if there are extra compact dimensions transverse to all stacks of branes then the string scale can be lowered to the TeV region  and the gauge hierarchy problem is solved even without the presence of N=1 supersymmetry \cite{extra}. The only known realization of this scenario where the Standard model
can be constructed in a string construction has appeared in the Standard-like models of \cite{Cremades:2002dh}, \cite{Kokorelis:2002qi} where the D5-branes are filling M4 and wrapping two-cycles in the extra dimensional space $T^4 \times C /Z_N$.
In \cite{Kokorelis:2002qi} we have generalized the 4-stack constructions of \cite{Cremades:2002dh} incorporating other four stack quiver SM's and also accommodating the deformed Standard model configurations that have been used in the 5-and 6-stack SM-like models of \cite{Kokorelis:2002zz}, \cite{Kokorelis:2002wa}.
In those constructions, the D5-branes get localized at the orbifold singularity and
only the massless chiral spectrum of Standard Model
remains at low energy (in addition to some scalars where in cases could become tachyonic).

    For the purposes of this paper we will silently accept, for the remaining of this section,
    that the relation (\ref{esti}) also holds and in SU(5)-like constructions from
    models with intersecting D5-branes, wrapping 2-cycles in $T^4 \times C / Z_N$ and
    having two extra dimensions transverse to the branes.
    String SU(5) models with two extra dimensions that localized the SM do not exist at present in the literature and may appear in \cite{kokosnow}.
In this way, if the $1 \ TeV \ \leq \ M_s \ < \ 10$ TeV we see that in case III, we get that the mass of the U-quark gets calculated in eqn.(\ref{esti10}).
\begin{align}
  m_U \stackrel{Low \ M_s}{\sim} \left\{  \begin{array}{ll}
         \mbox{$ M_w^4 \  M_s^{-3}\cdot e^{- {\tilde A}}$  GeV};\ \   \mbox{$\langle {\bf 5 }^H \rangle \sim \upsilon, \   \langle {\bf 1 }\rangle  \sim M_s \sim 1 \ TeV$,}&
         \makebox{Case  III }\\\nonumber &\\\nonumber
         \mbox{$M_s \cdot  \ e^{-{\tilde A}} \ GeV;$} \ \ \mbox{$\langle {\bf 5 }^H \rangle \sim \langle {\bf 1 }^H \rangle \sim M_s $ ,}&\makebox{Case  IV }
         \\\nonumber\end{array}
          \right .
          \end{align}
\beq
\label{esti10}
\eeq

\begin{table}[ht]
\centering
\begin{tabular}{||c||c|c|c|c|c|c|c|c|c|c||}
\hline\hline
${\bf M_s}$ & $1$ \ TeV &  $5$ \ TeV & $10$\ TeV  & $1$  \ TeV&  $5$ \ TeV& $9$ \ TeV & $10$ \ MeV\\\hline
${\bf M_u}$ &  $1.5$ \ MeV & $1.5$ \ MeV   & $1.5$ \ MeV  & $4$ \ MeV&  $4$ \ MeV & $4$ \ MeV & $4$ \ MeV \\\hline
${\bf A}$ &  $7.80$ & $2.97$   & $0.89$ & $6.82$ &  $1.99$ & $0.228$ & $-0.088$\\\hline
\hline\hline
\end{tabular}
\caption{Case III: Values of the String scale $M_s$ versus the area $A$, the mass of up(U)-quark $M_u$ and the vev of the singlet Higgses while
keeping constant the values of the up-quark and the vev of the electroweak Higgs of
 the order of the eletroweak scale $\upsilon =246$ GeV.
\label{tablegut0} } 
\end{table}
\underline{\em  In Theories with Extra dimensions: M$_s$ subject to $ 1 \ TeV  <  M_s \leq 9  $ TeV }
\newline
\newline
In particular the area factor expressed in terms of $M_s$, $M_u$, $M_w^4$ reads
\begin{align}
A = - \frac{M_s^3 \ M_U}{M_W^4 }
\end{align}

From table (\ref{tablegut0}), we observe that in Case III, when {\em the string scale is
low} and the Higgs $5^H$ takes a vev of the order of the electroweak scale $\upsilon =246$ GeV, {\em the string
scale is subject to an upper limit; its value can take any value from one (1) TeV to nine (9) TeV }. For $M_s = 10$ TeV the area turns negative(we kept constant the value of the U-quark mass). On the other hand in Case IV, there is no constraint in the value of the string scale from first principles as
the hierarchical mass of the up-quark 1.5-4.0 MeV is
derived from the area suppression in the exponential.

\section{Perturbative application to a 4D $\IZ_2 \times \IZ_2$  IIA N=1 SU(5) with D6-branes}

An example of a N=1 supersymmetric SU(5) model exhibiting the relevant perturbative mass term for the up-quarks can be seen by choosing for example the model II.1.1 of \cite{Cvetic:2002pj}. The models come from an orientifold compactification of type IIA on a six-dimensional torus in a background of a $\IZ_2 \times \IZ_2$ orbifold symmetry. The relevant rules for calculating the spectrum are summarized
in appendix A.
In this SU(5) model there are no filled branes to satisfy RR tadpoles. The wrappings are reproduced for convenience in table (\ref{tabgu}).
The initial gauge group is a $U(5) \times U(1) \times U(1)$.
\begin{table}[ht]
\centering
\begin{tabular}{|c||c|c||c|c|c|c|c|c|}
\hline
     & \multicolumn{8}{c|}{$U(5)\times U(1)\times U(1)$} \\
    \hline\hline \rm{D6} & $N$ & $(n^1,l^1)\times (n^2,l^2)\times
(n^3,l^3)$ & $n_{\Ysymm}$ & $n_{\Yasymm}$  & $b$ & $c$ & $b'$ & $c'$ \\
\hline
    $a$&  10& $(0,-1)\times (1,4)\times (1,1)$ & -3 & 3 & 0 & 24  & 0 & 0\\
    $b$&   2& $(-1,3)\times (-1,4)\times (1,1)$ & -6 & -42&- & 0  & - & -96 \\
    $c$&   2& $(-1,0)\times (-1,4)\times (7,1)$ & 27  & -27  &- &- & -& -\\
\hline
     & & \multicolumn{7}{c|}{N=1 / SUSY / conditions $x_B=4x_A=4x_C$}\\
\hline
\end{tabular}
\caption{N=1 $\IZ_2 \times \IZ_2$ $SU(5)$ chiral spectrum}
\label{tabgu}
\end{table}
There are two massless U(1)'s given by $6 U(1)_a + 5 U(1)_b$, $F_c$ and a massive $-5 U(1)_a + 6 U(1)_b $. The $F_c$ U(1) could be broken by the vev of the $1_{(0, 0, -2)}$ multiplet.

In the present case the up-quark masses
may receive perturbative contributions from the gauge invariant expression (\ref{newco}). For this particular model there are two different Yukawa contributing
to the mass of the up-quarks. In the first mass term, see eqn.(\ref{one1}), we are
using the Higgs 5-plets from the
N=2 ab sector and the gauge singlets from the bb$^{\prime}$ sector. In the second mass term contribution, see eqn. (\ref{koa3}), we are
using the Higgs 5-plets from the
N=2 ac$^{\prime}$ sector and the gauge singlets from the cc$^{\prime}$ sector.
\beq
 \frac{1}{M_s^5}{\bf 10}_{(2, 0, 0)} \cdot {\bf 10}_{(2, 0, 0)} \cdot {\bf {\bar 5}}^H_{(-1, 1, 0)} \cdot {\bf {\bar 5}}^H_{(-1, 1, 0)}  \cdot {\bf {\bar 5}}^H_{(-1, 1, 0)} \cdot {\bf {\bar 5}}^H_{(-1, 1, 0)} \cdot 1_{(0, -2, 0)}^H \cdot  1_{(0, -2, 0)}^H
 \label{one1}
 \eeq
 \beq
 \frac{1}{M_s^5}{\bf 10}_{(2, 0, 0)} \cdot {\bf 10}_{(2, 0, 0)} \cdot {\bf {\bar 5}}^H_{(-1, 0, -1)} \cdot {\bf {\bar 5}}^H_{(-1, 0, -1)}  \cdot {\bf {\bar 5}}^H_{(-1, 0, -1)} \cdot {\bf {\bar 5}}^H_{(-1, 0, -1)} \cdot 1_{(0, 0, 2)}^H \cdot  1_{(0, 0, 2)}^H
\label{koa3}
\eeq
The mass of the up-quarks can be distinguished according to the order of the
vacuum expectation value of the Higgs 5-plets and the gauge singlets. The
up-quark mass estimate to both cases of eqn's (\ref{one1}), (\ref{koa3}) is identical to that of eqn. (\ref{esti}).
See the subsection (4.2), for a comment on the associated Yukawa couplings of
eqn's (\ref{one1}), (\ref{koa3}), that can generate
additional contributions to the mass of the up-quarks through multi-instantons.
In the present class of models, the perturbative mass of the up-quarks is a linear combination of the couplings (\ref{one1}) and (\ref{koa3}).

\section{Instantons and missing SU(5) masses}

\subsection{General SU(5) u-mass instanton solutions}
An instanton mechanism  that
can potentially generate the missing up(down) quark coupling for an SU(5)/flipped SU(5) GUT of eqn. (\ref{as1}) was proposed
\cite{Blumenhagen:2007zk} (without presenting details on a particular O(1) instanton solution), that uses
three disk
diagrams involving the $SU(5)_a$ and the $U(1)_b$ D6-branes to absorb the
relevant zero modes. In this case the violation of the U(1) charge that has to be absorbed by the instanton is the excess charge $(U(1)_a, U(1)_b)= (5,1)$.

In this work, we find that the up(down) quark masses
may also receive contributions to their masses via a different instanton mechanism.  It can be generated by an E2-instantons carrying gauge group O(1), Sp(2) or U(1) and having the appropriate zero mode structure. For the case of the gauge group O(1) we present in the next section the explicit form of the O(1) stringy instanton.
We find that stringy instantons give rise to the up(down) quark mass coupling
via the coupling
\beqa
\frac{1}{M_s^3}\langle {\bf 10}_{(2,0)} \cdot {\bf 10}_{(2,0)} \cdot
{\bf{\bar 5}}^H_{(-1, -1)} {\bf{\bar 5}}^H_{(-1, -1)}  {\bf{\bar 5}}^H_{(-1, -1)}  {\bf{\bar 5}}^H_{(-1, -1)} \rangle \  , \\
\label{coupli}
\eeqa
where we have used the general two stack D6-brane SU(5) Grand Unified model structure of section II. The excess charge which has to be absorbed by the instanton
is $U(1)_b = -4$.
This coupling violates the $U(1)_b$ symmetry and thus is perturbatively forbidden.
Hence a non-perturbative contribution to the superpotential generated by a E2 instanton is
a viable possibility.
A general E2 instanton wrapping a three cycle and placed away from the orientifold fixed plane gives rise to four
bosonic zero modes arising from the breakdown of the Poincare invariance and also four
fermionic zero modes $\theta_a$, ${\overline{\tau}}_{\dot{\alpha}}$ in the uncharged sector.

 To generate a superpotential contribution we need instantons
with {\em O}(1) Chan-Paton symmetry. These instantons  are wrapping a rigid orientifold invariant three cycle, generate an
{\em O}(1) gauge group on their volume and possess the two $\theta_a$ uncharged instanton zero modes which saturate the $d^2 \theta$ 4d superspace integration.
Hence each rigid O(1) instanton carries two neutral fermion zero modes, the
goldstinos of the susys it breaks.
However,
as was noted in \cite{isu} instantons with
Sp(2) or U(1) CP symmetries may also generate the relevant superpotentials if there is additional dynamics, e.g. fluxes, that can saturate the extra zero modes that could appear in those cases.
 Hence the induced instanton charge that cancels the $U(1)_b$ violation
has the following intersection numbers between the $E_2$-instanton and the D6 b-brane
\beqa
Sp(2) & case & \ : \ \ I_{E_2 b} = -2, \ \nonumber\\
{\cal O}(1) & case &  \ : \ \ I_{E_2 b} = -4, \   \nonumber\\
U(1) & case & \ : \ \ I_{E_2 b} = -2, \ I_{E_2 b^{\prime}} = 2 \ . \nonumber\\
\label{inter}
\eeqa
Thus in general we need four charged zero modes, a "doublet"\footnote{The term "doublet" is used in a broad sense; not connected to the group representation.}of $\alpha^i_b$ zero modes coming
from the intersection of the instanton M and the $D6_b$ brane and another "doublet" of
$\gamma^i$ zero modes coming the intersection of M with $D6_{b^{\prime}}$.
Since the instanton lies in a $\Omega{\bar \sigma}$ invariant position this guarantees that the uncharged part of the instanton measure contains two fermionic
$\theta_a $ and four bosonic degrees of freedom $x^{\mu}$.
The zero modes appearing in the intersections (\ref{inter}) gets saturated by two identical disk diagrams seen in figure 1.

\begin{figure}[h]
\vskip 1cm
\begin{center}
\includegraphics[width=0.8\textwidth]{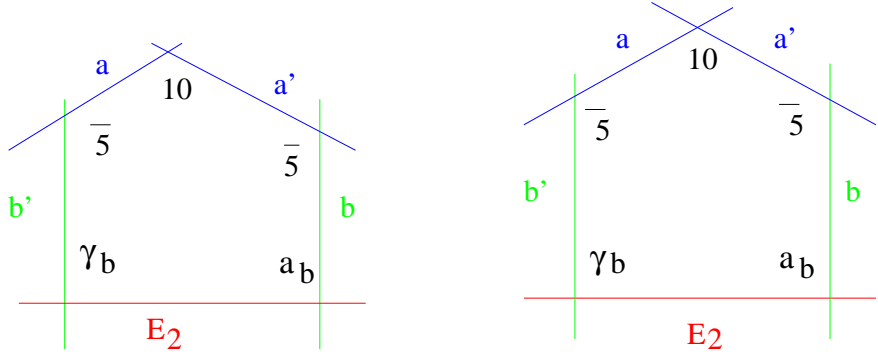}
\end{center}
\caption{\small Absorption of charged zero modes by the instanton}\label{yukafig}
\end{figure}

These disk diagrams induce the non-perturbative contribution to the Yukawa coupling
for the u-quarks which is based on the calculation of the path integral
\beqa
\frac{1}{M_s^3}\int d^4 x \ d^2 \theta \ d^2 \alpha \ d^2 \gamma e^{-S^{clas}_{E_2}} e^{Z^{\prime}}
\langle \alpha  X \gamma  \rangle \langle \alpha X \gamma \rangle \ , \nonumber\\
 X_i = X_j = {\bf 10} \cdot {\bf \overline{5}}^H \cdot  {\bf \overline{5}}^H \ .
 \label{insa}
\eeqa
 Also $S^{clas}_{E_2}$ is the classical instanton action for the $E_2$ instanton, $e^{Z^{\prime}}$ is the holomorphic part of the one-loop determinant arising from the
annulus and M$\ddot{o}$bius diagrams ending on the instanton and the D6-branes or O-plane respectively; it can be interpreted as the one-loop Plaffian.
The existence of the path integral denotes the fact that all charged zero modes arising at the instanton intersections with D6-branes can be soaked up via
disk diagrams $ <  \lambda^a_{-1/2} \Phi_{ab} \lambda^b_{-1/2} > $ , where $\Phi_{ab} =  \phi_{ab} + \theta \psi_{ab}$ denotes the chiral superfield arising at the intersection of branes a and b;
$\lambda_i$ denote the charged zero modes (see for example \cite{Cvetic:2007sj}).
The instanton suppression factor in (\ref{insa}) is
\begin{align}
e^{-S^{cl}_{E2_1}}=e^{-\frac{2 \pi}{l^3_s\, g_s} \,
\text{Vol}_{E2_1}} =e^{-\frac{2\pi}{\alpha_a}\,
\frac{\text{Vol}_{E2_1}}{\text{Vol}_{D6_a}}}\,\,,
\label{suppression1}
\end{align}
where the $\frac{\text{Vol}_{E2_1}}{\text{Vol}_{D6_a}}$ is
given by
\begin{align}
\frac{\text{Vol}_{E2_1}}{\text{Vol}_{D6_a}}=
\frac{1}{2}\left(\prod_I \left[ \frac{(n^I_{E2_1})^2 + (\widetilde
m^I_{E2_1})^2 U_I^2}{(n^I_{a})^2 + (\widetilde m^I_{a})^2
U_I^2}\right]\right) ^{1/2}\,\,. \label{supp2}
\end{align}
To generate the $\cal O$(1) MeV order of the up-quarks, or the $\cal O$(200) GeV mass
of the top quark, assuming that there is no perturbative term which contributes to the quark masses,
we need instanton suppression \footnote{Additional worldsheet instanton suppression
factors could also be created in the calculation of disk amplitudes associated with the 
intersection structure of the participating branes \cite{Cvetic:2007ku}.} factors
\begin{align}
e^{-S_{E2_1}^{cl}} \stackrel{u-quark}{\sim} 10^{-21} \ , \ \
e^{-S_{E2_1}^{cl}} \stackrel{t-quark}{\sim} 10^{-16}
\label{sup1}
\end{align}
respectively, which are rather large estimates. We assume that the string scale $\approx 10^{18}$ GeV. Of course models with large numbers of exotics could achieve that goal, as we comment in the next section. Clearly such a model has to be presented for this scenario to be realized. See next section, where in a particular local N=1 $\IZ_2 \times \IZ_2^{\prime}$ model, typical suppression
factors are in the range $10^{-4}$ for the MSSM and can be as large as $10^{-21}$ when large number of exotics are present. Thus in general the instanton eq.( \ref{insa}),
constitutes a subleading correction to the perturbative mass term of the U-quarks, the latter given in eq.(\ref{koa1}). \newline 
To be more concrete lets us comment on the 3 x 3 Yukawa coupling mass matrix for the up-quarks. The full matrix is a sum of a perturbative and a non-perturbative part.
The perturbative generated Yukawa contribution to the masses, expressed 
by eq.(\ref{koa1}), 
leads to Yukawa coupling matrices in the form
\begin{align}
Y^{Per}_{u^I}=\left(\begin{array}{ccc} A^u_{11}&A^u_{12}&
A^u_{13}\\
A^u_{21}&A^u_{22}& A^u_{23}\\
A^u_{31}&A^u_{32}&A^u_{33}\\
\end{array}\right)
\end{align}
while the non-perturbatively generated instanton part is also given by
\begin{align}
 Y^{NP}_{u^I }=\left(\begin{array}{ccc}
B^u_{11}&B^u_{12}&
B^u_{13}\\
B^u_{21}&B^u_{22}& B^u_{23}\\
B^u_{31}&B^u_{32}& B^u_{33}\\
\end{array}\right).
\end{align}
We note that in general due to the instanton suppression factor
$A_{ij}\gg B_{kl}$ may hold for arbitrary $i,j,k,l$. 
The precise form of the $A_{ij}$ also depends on the intersection numbers of the 
particular D-brane model. 
For the
class of O(1) instantons in eqn.(\ref{insa}) the mass matrix factorizes, giving a rank
one matrix, thus the instanton contributes a correction to only one generation
of U-quarks. As there is present the perturbative contribution to the U- mass,
the instanton correction suggest to us that it may be contributing to the heaviest generation,
the third (assuming that it is the one associated with the top quark). A different scenario could be realized in the following way. 
Let us suppose that the perturbative mass matrix has a texture that it leads to
a rank one matrix.
In this case, we can identify the non-zero perturbative mass eigenvalue with the mass of the third generation generating the mass of the top quark, while the rank one instanton
mass matrix can contribute to the mass e.g. of the second heaviest generation of the up-quarks, leaving massless the lightest generation, the first one.

After integrating out the charged zero modes the superpotential instanton contribution appears
\begin{align}
\int d^4 x d^2 \theta \ Y^{\prime}_{U(D)}\ 10 \ 10 \ { \overline{5}}\ { \overline{5}} \ { \overline{5}}\ { \overline{5}} \ .
\end{align}
The factor $Y$ accommodates the classical suppression factor and $e^{Z^{\prime}}$, as
well as the contribution of the disk amplitude $\langle \alpha  X \gamma  \rangle \langle \alpha X \gamma \rangle$, the latter depending on the open string moduli in the sums of worldsheet instantons that connect the intersection points.

\subsection{Instantons on the N=1 local model of section 3}

Lets us now make some observations concerning the N=1 $\IZ_2 \times \IZ_2$ string model appearing in section 3. We have seen that for this model there are two different mass terms contributing to the mass of the up-quarks. Accordingly, we assume that
the wrappings of this model constitute the "observable part' of a local model on a
 N=1 $\IZ_2 \times \IZ_2^{\prime}$, with no tadpole cancelation at this level.
In this case, we find that there
are two potential contributions to the up-quark mass through instantons. To
saturate the instanton zero modes in this case, we will need a duplicate of the
two diagrams of figure 1; the one pair identical to that of figure 1 and the
second pair having replaced the b D6-brane with the c D6-brane.
The instanton couplings for which instantons that may also additionally contribute corrections to the u-masses are
 \beq
 \frac{1}{M_s^5}{\bf 10}_{(2, 0, 0)} \cdot {\bf 10}_{(2, 0, 0)} \cdot {\bf {\bar 5}}^H_{(-1, 1, 0)} \cdot {\bf {\bar 5}}^H_{(-1, 1, 0)}  \cdot {\bf {\bar 5}}^H_{(-1, 1, 0)} \cdot {\bf {\bar 5}}^H_{(-1, 1, 0)}
 \label{one11}
 \eeq
with excess charge $U(1)_b = 4$ to be absorbed by one instanton and also
\beq
 \frac{1}{M_s^5}{\bf 10}_{(2, 0, 0)} \cdot {\bf 10}_{(2, 0, 0)} \cdot {\bf {\bar 5}}^H_{(-1, 0, -1)} \cdot {\bf {\bar 5}}^H_{(-1, 0, -1)}  \cdot {\bf {\bar 5}}^H_{(-1, 0, -1)} \cdot {\bf {\bar 5}}^H_{(-1, 0, -1)}
\label{koa33}
\eeq
with excess charge to be "absorbed" by another instanton $U(1)_c = -4$.
In this case, we will need at least two different O(1) instantons to absorb the relevant charges. An example of a single O(1) instanton compensating an excess charge -4 and generating the up-quark masses
is described in the next section.
In principle a set of two instantons typically has more neutral zero modes
than is needed to lead to a superpotential.
If you consider a set of two O(1)
instantons, in principle you have eight charged fermion zero modes that have to be absorbed.
For a multi-instanton process to contribute to the superpotential, one
needs to make sure that all the extra fermion zero modes are lifted by
some interaction. This has been shown only in few examples in \cite{GarciaEtxebarria:2008pi}.
More details will be presented elsewhere.

\subsection{O(1) instantons for U-masses to a local $\IZ_2 \times {\IZ_2}^{\prime}$  IIA N=1 SU(5)}

In this section, we will examine local setups of the stringy O(1) instanton, that
can generate perturbatively the up-quark mass in SU(5) models, based on a background
of an orientifold $T^6/\IZ_2 \times \IZ_2^{\prime}$ with Hodge numbers $(h_{11}, h_{22})=
(3, 51)$, the so called $T^6/\IZ_2 \times \IZ_2^{\prime}$ with discrete torsion. In these models, the orbifold quotient and the choice of discrete torsion is such that
they contain rigid branes that freeze the position of the branes at the fixed points
of the orbifold fixed points where the twisted cycles arise. Hence the corresponding adjoints associated to the transverse translations of the branes may not exist
(models with completely rigid cycles have been considered in \cite{Larosa:2003mz}. See also \cite{Dudas:2005jx} for related constructions in type IIB).

Lets us consider the chiral spectrum displayed in table (\ref{sp01}). It corresponds
to an initial $SU(5)_a \times U(1)_a \times U(1)_b$ gauge group at the string scale.
\begin{table}[ht]
\centering
\begin{tabular}{|c|c|c|}
\hline
sector & $|I_{ij}|$ &  $SU(5)_a \times U(1)_a \times U(1)_b$   \\
\hline \hline
$(a,a')$ &  $3$ & ${\bf {{{10}}}}_{(2, 0)}$ \\
 $(a,a')$ &  $1$ & ${\bf \overline{15}}_{(-2, 0)} $     \\
$(a,b)$ &  $38$ & ${\bf \overline{5}}_{(-1, 1)}$    \\
$(a,b')$ &  $0$ & ${\bf 5}_{(1, 1)}^H + {\bf \overline{5}}_{(-1, -1)}^H$    \\
$(b,b')$ &  $26$ & ${\bf 1}_{(0, 2)}$  \\
\hline
\end{tabular}
\caption{Chiral matter for local N=1 $SU(5)$ GUT model with intersecting D6-branes
on the $\IZ_2\times \IZ_2^{\prime}$ orientifold. }
\label{sp01}  
\end{table}
The mass of the up-quarks is given by the coupling
${10}_{(2, 0)} \cdot {10}_{(2, 0)} \cdot 5_{(1, 1)}^H$ which is not allowed in perturbation theory due to the non-conservation of $U(1)_b$ charge.
We will rather generate this coupling in this section in two ways, a) via non-perturbative effects from O(1) instantons and b) from perturbation theory at the end of the section.

The matter D6-branes are described by fractional branes that carry charge under one twisted sector. They wrap D6-branes along untwisted bulk and twisted three
cycles. The bulk part of the wrappings is given in table (\ref{tabgu10}).
\begin{table}[ht]
\centering
\begin{tabular}{|c|c|c|c|}
    \hline\hline brane & $(n^1,l^1)$& $(n^2,l^2)$& $(n^3,l^3)$ \\
\hline
    $N_a = 5$ &  $(3, 4)$ & $(0,-1)$ & $(1,1)$ \\\hline
    $N_ b = 1$ &  $(1, 3)$ & $(-4,1)$ &$(-1,1)$ \\
\hline
\hline
\end{tabular}
\caption{N=1 $Z_2 \times Z_2^{\prime}$ $SU(5)$ Bulk wrapping numbers}
\label{tabgu10}
\end{table}
For the choice
\begin{align}
U^1 U^3 = \frac{3}{4},\ \, \, -12U^3 + 36U^1-13 U^1U^2U^3=0 ,
\label{susy}
\end{align}
 N=1 supersymmetry is obeyed by the D6-branes, as they align with the orientifold planes. The complex structure moduli $U_{i} = (R^2/R_1)^{(i)}$, where i denotes the
 i-tori could be fixed according to the eqn's (\ref{susy}). An example value, solving the complex srtucture constraints (\ref{susy}), is the choice
 \begin{align}
 U^1 = 3 \ , U^2 = \frac{140}{13}  \ , U^3 = \frac{1}{4}
 \end{align}
For definiteness we make the choice of crosscap orientifold charges
\begin{align}
\eta_{\Omega {\cal R}}=1 \qquad  \eta_{\Omega {\cal R} \theta}=1
\qquad  \eta_{\Omega {\cal R} \theta'}=-1 \qquad \eta_{\Omega {\cal
R} \theta\theta'}=1
\label{orienti1}
\end{align}
The fractional matter D6-branes wrap fractional cycles, cycles
are charged only under one twisted sector $g$ in the form
\begin{align}
 \pi^F= \frac{1}{2} \pi^B +
\frac{1}{2} \left( \sum_{i,j \in S_{g}} \epsilon^{g}_{ij}
  \Big[\alpha^{g}_{ij}\times (n^{I_g}_a,\widetilde{m}^{I_g}_a)
  \Big]\right)\,\,.
  \label{interfra}
\end{align}
This class of cycles are only rigid in two tori and can move freely
in the torus invariant under the action $g$. Then, for two
fractional branes charged under a different twisted sector
 the intersection number is simply given by
 \begin{align}
\pi^F_a \circ \pi^F_b =\prod^3_{i=1} (n^i_a \widetilde{m}^i_b-n^i_b
\widetilde{m}^i_a)
\label{interbra}
\end{align}
The fractional \footnote{Further details on the notation we are using
can be seen in appendix B. We are using an identical notation to \cite{Ibanez:2008my}.} D6-branes generating the chiral matter spectrum of table (\ref{sp01})  read
\begin{align}
a:& \ \frac{1}{2}[(3, 4)(0, -1)(1, 1)] + \,\,\,\,\frac{1}{2}\left( \sum_{i,j \in (2,4)\times (3, 4)}[ a_{ij}^{\th} \times (1, 1)]\,\,\,\,\,\,\,\right)\nonumber\\
b:& \ \frac{1}{2}[(1, 3)(-4, 1)(-1, 1)] + \frac{1}{2}\left( \sum_{i,j \in (1,4)\times (3, 4) }[ a_{ij}^{\th } \times (-1, 1)]\right)
\end{align}
Given our choice of fractional branes, if we were making another choice of crosscap charges e.g. as the one in appendix 8 of \cite{Blumenhagen:2005tn}, our new choice may only have affected the numbers of antisymmetric and symmetric representations in the spectrum. In the latter case all K-theory constraints listed in \cite{Blumenhagen:2005tn} would have been satisfied.
The next, beyond the tree level, general gauge invariant responsible for generating
the mass of the up-quarks in the local SU(5) model via O(1) instantons is
\begin{align}
\frac{1}{M_s^3}\langle {\bf {10}}_{(2,0)} \cdot {\bf {10}}_{(2, 0)} \cdot
{\bf \overline{5}}^H_{(-1, -1)} \cdot {\bf \overline{5}}^H_{(-1, -1)} \cdot {\bf \overline{5}}^H_{(-1, -1)} \cdot {\bf \overline{5}}^H_{(-1, -1)}  \rangle
\label{non-pe1}
\end{align}
This mass coupling is not allowed in perturbation theory as there is a non-canceled $U(1)_b = -4$ excess charge.
We have used the N=2 Higgses ${\bar 5}$-plets from the ab$^{\prime}$ sector in order to generate it.
The O(1) instanton able to  and generating the dim=7 operator (in superpotential form) 
\begin{align}
W_{non-pert} = \frac{1}{M_s^4}
{\bf {10}}_{(2,0)} \cdot {\bf {10}}_{(2, 0)} \cdot
{\bf \overline{5}}^H_{(-1, -1)} \cdot {\bf \overline{5}}^H_{(-1, -1)} \cdot {\bf \overline{5}}^H_{(-1, -1)} \cdot {\bf \overline{5}}^H_{(-1, -1)}
e^{-S_{ins}}
\label{insta}
 \end{align}
thus absorbing the excess zero modes and giving a mass to the up-quarks is given by
\begin{align}
E2_1 =& \frac{1}{4}[(1, 0)(0, 1)(0, -1)] + \frac{1}{4}\left( \sum_{i,j \in (1,3)\times (3, 4) }[ a_{ij}^{\th} \times (0, -1)\right)&\nonumber\\
 \,\,\,\,\,\,+& \frac{1}{4}\left( \sum_{i,j \in (\star, \star) \times (\star, \star) }
 [ a_{ij}^{ {\th}^{\prime} } \times (1, 0)] \right)+ \frac{1}{4}\left( \sum_{i,j \in  (\bullet, \bullet) \times (\bullet, \bullet)}[ a_{ij}^{\th \th^{\prime}} \times (0, 1)]\,\,\,\right) .&
 \label{instan1}
\end{align}
The fractional instanton is charged under all the different twisted sectors of the orbifold.
The meaning of $(\star,\star) \times (\star, \star)$ is that we can choose any combination of the following set of fixed points $\{ (1,2) \times (1, 2); \ (1,2) \times (3, 4); \ (3,4) \times (3, 4); \ (3,4)\times (1, 2) \}$ the bulk wrappings of the fractional brane can go through, as there is no restriction from first principles. In a similar way\footnote{we follow the notation of the fixed points detailed in \cite{Blumenhagen:2005tn}.}, $(\bullet, \bullet) \times (\bullet, \bullet)$ means that we can choose any combination of the following sets $\{ (1,3) \times (1, 2); \ (1,3) \times (3, 4); \ (2,4) \times (1, 2); \ (2,4)\times (3, 4) \}$.
The instanton of eqn.(\ref{instan1}) gives rise to the intersection
number pattern
\beq
I_{E2_{1} a} = 0,\ \ I_{E2_{1} b} = -4
\label{internum}
\eeq
thus canceling the $U(1)_b$ charge violation of (\ref{non-pe1}). The relevant string
diagrams generating the superpotential term are given in figure (1).
All the states from the N=2 sector $E_{2_{1}} - a$ are massive. The bulk part of both $E_{2_{1}}$, $a$, D6-branes is parallel across the second torus, while simultaneously the twisted part of the branes runs through different fixed points. Thus all N=2 states gets massive and
there are no vector-like states from this sector.
The factor $\frac{1}{2}$ is due to the fact that the $D6$-brane
wraps a fractional cycle while the instanton a rigid one.

Let us now derive some estimates on the instanton suppression factor for the local models of this section.
Let us assume $\alpha_a=1/24$, as in the MSSM, at string
scale. If we choose the complex structure moduli, from eqn.(\ref{susy}), to be stabilized by the choice $U^1 =1$, a N=1 consistent solution for the complex structure is $U^3 =3/4$, $U^2 = 36/13$. In this way,
we get the suppression factor
\begin{align}
e^{-S^{cl}_{E2_1}} \sim \ 10^{-4} \ .
\end{align}

Several values of $a_{GUT}$ against the suppression factor may be seen in table (\ref{supr1}). Cases with the mark $\times :$ are phenomenologically excluded
as their contribution is bigger that the order of the mass of the quark.

\begin{table}[ht]
\centering
\begin{tabular}{|c|c|c|c|c|c|}
    \hline\hline $a_{GUT}$ & $1/24$& $1/30$  & $ $1/123&  $1/124$ & $125$\\
\hline
$e^{-S^{cl}_{E2_1}}$  &  $10^{-4}$ & $10^{-5}$   & $4.43 \times 10^{-21}$ & $3.03 \times 10^{-21}$& $2.07 \times 10^{-21}$\\
  \hline
  $M_u$ (MeV)  & $\times : \ 10^{17}$  & $\times : \ 10^{16}$   & $4.43$ & $3.03$ & $2.07$ \\
\hline
\hline
\end{tabular}
\caption{$a_{GUT}$ and instanton suppression factor values against the instanton  contribution to the U-quark masses, for Georgi-Glashow SU(5)-type GUTS, from eqn.(\ref{massla}).}
\label{supr1}
\end{table}
We observe that for D-brane models with a lot of extra matter $a_{GUT} >>$, instanton
suppression factors become large enough, so that they contribute significant corrections to the mass of the U-quarks for Georgi-Glashow GUTS( Equivalently for the instanton contribution to the down quarks for Flipped SU(5) GUTS).

Apart from string instantons which give rise to such a superpotential
mass term in eqn. (\ref{insta}), a perturbative contribution to the up-quark mass exists for the present $\IZ_2 \times \IZ_2^{\prime}$ models in the form
\beq
M_{pert}^{U} =  \frac{10_{(2, 0)}  10_{(2,0)}}{M_s^5}
\langle {\bf 5}^H_{(-1, -1)} \rangle\cdot
\langle {\bf 5}^H_{(-1, -1)} \rangle\cdot
\langle {\bf 5}^H_{(-1, -1)} \rangle\cdot
\langle {\bf 5}^H_{(-1, -1)} \rangle\cdot
\langle 1_{(0, 2)} \rangle
\cdot \langle 1_{(0, 2)} \rangle e^{-A}
\label{massla}
\eeq
giving a mass of order $\approx M_s e^{-A}$.

We have shown that the long standing problem of missing up(down) quark masses
that has singled out the global SU(5)/flipped SU(5) D-brane models in orientifold
compactifications of type II theory can be resolved in terms of the previously unknown Grand Unified SU(5) GUT invariants
\begin{align}
{\bf 10} \cdot {\bf 10} \cdot {\bf \overline{5} }^H \cdot {\bf \overline{5}}^H \cdot {\bf \overline{5} }^H  \cdot {\bf \overline{5} }^H
\label{newinv}
\end{align}
from higher orders of perturbation theory.
Hence in all SU(5) models in IIA theory from intersecting branes, or in its dual theory in 4D SU(5)'s from IIB with or without fluxes, the relevant up-quark masses gets generated, minimally, from the dim=9 mass terms (\ref{koa1}). As we have noted a perturbative mass term is not possible for flipped SU(5) GUTS.

We have also shown that Euclidean E2-instantons based on the new couplings
can also contribute NP corrections to the mass of the relevant quarks.
The presented instanton related mass corrections arising in eqn.(\ref{newinv}) from SU(5) gauge invariants are also valid for the flipped SU(5) GUTS.

These corrections can be also extended
 to perturbatively generate missing up-quark and also its instanton associated u-quark mass contributing invariants,
  in higher GUTS. These are the ones with SU(6)/flipped SU(6) gauge groups where the problem of missing up(down)-quark masses is more acute as also the lepton masses are not allowed in perturbation theory. The relevant gauge invariants now appear in the
 form of
 \begin{align}
{\bf 15} \cdot {\bf 15} \cdot {\bf \overline{6} }^H \cdot {\bf \overline{6}}^H \cdot {\bf \overline{6} }^H  \cdot {\bf \overline{6} }^H \ ,
  \label{newinv6}
\end{align}
 in terms of SU(6) representations and are discussed in \cite{kokosgut6}.

\section{Acknowledgements}
We thank  L. Ibanez, F. Marchesano, R. Richter and A. Uranga for useful discussions.


\section{Appendix B - $T^6/{\mathbb Z}_2 \times {\mathbb Z}_2$ Orientifold \label{appendix orientifold1} }

In this section we summarize the main
points of the construction, and the details of our notation we are using in section 3.

We start with type IIA theory on $\IT^6/(\IZ_2 \times
\IZ_2)$; with the orbifold group generators $\theta$, $\omega$  acting on the
complexified coordinates of a factorizable  $\IT^6 = \IT^2 \times \IT^2 \times \IT^2$ tori as
\beq
 \theta: \ (z_1,z_2,z_3) \to (-z_1,-z_2,z_3); \
 \theta': \ (z_1,z_2,z_3) \to (z_1,-z_2,-z_3).
 \label{comp}
 \eeq

We implement an orientifold projection by $\Omega R$, where
$\Omega$ is world-sheet parity, and $R$ acts as \beqa R:
(z_1,z_2,z_3) \to ({\ov z}_1,{\ov z}_2,{\ov z}_3). \eeqa
To cancel the RR charge of the O6-planes, we introduce D6-branes
wrapped on three-cycles that are products of one-cycles in each of
the three two-tori.
Each one-cycle is described by the wrapping numbers $(n_a^i, l_a^i)$.
Under an $\Omega R$ reflection a cycle $(n_a^i, l_a^i)$ is mapped to
$(n_a^i,-l_a^i)$. That means that at the
level of the spectrum, for a stack of $N_a$ D6-branes along the cycle $(n_a^i,l_a^i)$ we also need to include its
image with wrapping numbers $(n_a^i,-l_a^i)$.
A torus can be tilted or untilted. For a tilted torus $l-n$ is even.
Avoiding multiply wrapped branes requires that $m$
and $n$ are relatively coprime, for tilted/untilted
tori.

By convention
to describe rectangular and tilted tori cycles we define \beq l_{a}^{i}\equiv
m_{a}^{i},\;{\rm rectangular, }\;\;\;\;\;\; l_{a}^{i}\equiv
2\tilde{m}_{a}^{i}=2m_{a}^{i}+n_{a}^{i},\; {\rm tilted}. \eeq.
 The intersection numbers of the
homology cycles are computed using
 \beq
\begin{array}{ll}
I_{ab}=[\Pi_a][\Pi_b]=2^{-k}\prod_{i=1}^3(n_a^il_b^i-n_b^il_a^i),&
I_{ab'}=[\Pi_a]\left[\Pi_{b'}\right]=-2^{-k}\prod_{i=1}^3(n_{a}^il_b^i+n_b^il_a^i)
\\\\
I_{aa'}=[\Pi_a]\left[\Pi_{a'}\right]=-2^{3-k}\prod_{i=1}^3(n_a^il_a^i),&
\\\\
\multicolumn{2}{l}{I_{aO6}=[\Pi_a][\Pi_{O6}]=2^{3-k}(-l_a^1l_a^2l_a^3+l_a^1n_a^2n_a^3+n_a^1l_a^2n_a^3+n_a^1n_a^2l_a^3)}
\end{array}
\label{intersections}\eeq where $k=\beta_1+\beta_2+\beta_3$ is the
total number of tilted tori; by definition $\beta = 0$ for untilted tori;
$\beta = 1$ for tilted tori.

The open string spectrum for branes intersecting at angles appears in \cite{hep-th/0107166} and is described in table (\ref{matter}).
\begin{table}
[htb] \footnotesize
\renewcommand{\arraystretch}{1.25}
\begin{center}
\begin{tabular}{|c|c|}
\hline
{\bf Sector} & {\bf Representation}
\phantom{more space inside this box} \\
\hline\hline
$aa$   & $U(N_a/2)$ vector multiplet  \phantom{}\\
       & 3 Adj. chiral multiplets  \phantom{} \\
\hline
$ab+ba$   & $I_{ab}$ $(\fund_a,\antifund_b)$ fermions  \phantom{} \\
\hline
$ab'+b'a$ & $I_{ab'}$ $(\fund_a,\fund_b)$ fermions \phantom{} \\
\hline $aa'+a'a$ & $\frac 12 (I_{aa'} - \frac 12 I_{a,O6})\;\;
\Ysymm\;\;$ fermions  \phantom{} \\
          & $\frac{ 1}{2} (I_{aa'} + \frac{1}{2} I_{a,O6}) \;\;
\Yasymm\;\;$ fermions \phantom{} \\
\hline
\end{tabular}
\end{center}
\caption{\small Spectrum of intersecting D6-branes
\label{matter} }
\end{table}


\section{Appendix C - $T^6/{\mathbb Z}_2 \times {\mathbb Z}_2'$ Orientifold }

In this appendix we briefly review the $T^6/{\mathbb Z}_2
\times {\mathbb Z}_2'$ orientifold with Hodge numbers
$(h_{11},h_{12})=(3,51)$ following
\cite{Blumenhagen:2005tn}, to which we refer the reader for further
details. The orbifold group acts as in eqn.(\ref{comp}) and where the combination $\theta\theta'$ acts as a
reflection in the first and third torus.
Also present are the bulk cycles
 \bea \Pi_a^B = 4 \, \bigotimes_{I=1}^3
\,(n_a^I [a^I] + \widetilde m_a^I [b^I]), \eea described in terms of
the fundamental one-cycles $[a^I], [b^I]$ of the $I$-th $T^2$ and
the associating wrapping numbers $n_a^I$ and $\widetilde m_a^I=
m_a^I + \beta^I n_a^I$. we note that $\beta^I$ takes the value $0$ and $1/2$
for rectangular and tilted tori, respectively.
We have also present the $g$-twisted cycles.  The action of the group elements, $\theta$,
$\theta'$ and $\theta\theta'$ possess 16 fixed points which after
blowing up give rise to two-cycles of ${\mathbb P}_1$. Together with the fundamental one-cycle invariant under
the respective group action the $g$-twisted cycles are constructed as \bea
\Pi^g_{ij} = \Big[\alpha^g_{ij} \times
(n^{I_g},\widetilde{m}^{I_g})\Big]. \eea In particular ${i,j} \in \{1,2,3,4\}
\times \{1,2,3,4\}$ labels one of the 16 blown-up fixed points of
the orbifold element $g  = \theta, \theta', \theta \theta' \in
\IZ_2\times \IZ_2'$; $I_g$ represents the $g$-invariant one-cycle with
$I_g = 3,1,2$ for $g  = \theta, \theta', \theta \theta'$.

Rigid cycles are charged under all three sectors $\theta$, $\theta'$
and $\theta\theta'$ and take the form \bea \Pi^F = \frac{1}{4} \Pi^B
+ \frac{1}{4} \Bigl( \sum_{i,j \in S_{\theta}}
\epsilon^{\theta}_{ij} \Pi^{\theta}_{ij} \Bigr)+ \frac{1}{4} \Bigl(
\sum_{j,k \in S_{\theta'}} \epsilon^{\theta'}_{jk}
\Pi^{\theta'}_{jk}  \Bigr) + \frac{1}{4} \Bigl(  \sum_{i,k \in
S_{\theta \theta'}} \epsilon^{\theta \theta'}_{ik} \Pi^{\theta
\theta'}_{ik}  \Bigr), \eea where $S_g$ is the set of fixed points
in the $g$-twisted sector. The $\epsilon^g_{ij}=\pm 1$ correspond to
the two different orientation the brane can wrap the blown up
${\mathbb P}_1$; they are subject to other consistency conditions which can be seen
in detail in \cite{Blumenhagen:2005tn}, including RR tadpoles and homological K-theory theory constraints. We do not list tadpole equations as for the
purpose of this work, we use only local models on the $\IZ_2 \times \IZ_2^{\prime}$.

The orientifold action $\Omega\mathcal{R}$ for the bulk cycles appears as
\begin{align}\label{orientifold action1}
\Omega\mathcal{R}:
[(n_1,\widetilde{m}_1)(n_2,\widetilde{m}_2)(n_3,\widetilde{m}_3)]\rightarrow
[(n_1,-\widetilde{m}_1)(n_2,-\widetilde{m}_2)(n_3,-\widetilde{m}_3)].
\end{align}
For the $g$-twisted cycle the $\Omega\mathcal{R}$ action is defined as
\begin{align}
\label{Omegatwisted} \Omega\mathcal{R}: \,\Big[\alpha^g_{ij}\times
(n^{I_g},\widetilde{m}^{I_g})\Big]\rightarrow -
\eta_{\Omega\mathcal{R}}\,\eta_{\Omega\mathcal{R}g} \Big[
\alpha^g_{\mathcal{R}(i)\mathcal{R}(j)}
\times(n^{I_g},-\widetilde{m}^{I_g})\Big],
\end{align}
where $\eta_{\Omega\mathcal{R}g}=\pm 1$ denotes the orientifold
charges of the different sectors and obeys
\begin{align}
\label{etaconst}
\eta_{\Omega\mathcal{R}}\,\eta_{\Omega\mathcal{R}\theta}\,
\eta_{\Omega\mathcal{R}\theta'}\,\eta_{\Omega\mathcal{R}\theta\theta'}=-1.
\end{align}
The reflection $\mathcal{R}$ leaves all fixed points of an untilted
two-torus invariant and acts on the fixed points in a tilted
two-torus as
\begin{align}
\mathcal{R}(1)=1, \qquad \mathcal{R}(2)=2, \qquad\mathcal{R}(3)=4,
\qquad\mathcal{R}(4)=3.
\end{align}
Under the orientifold action \eqref{orientifold action1} the orientifold
appears as
\begin{align*}\pi_{O6}=\,&
2^3\, \eta_{\Omega\mathcal{R}}\, [(1,0)(1,0)(1,0)]
+2^{3-2\beta_1-2\beta_2} \,\eta_{\Omega\mathcal{R}\theta}\,
[(0,1)(0,-1)(1,0)]\\ \\& +2^{3-2\beta_2-2\beta_3}
\,\eta_{\Omega\mathcal{R}\theta'}\, [(1,0)(0,1)(0,-1)]+ 2^{ 3-
2\beta_1 -2\beta_3} \,\eta_{\Omega\mathcal{R}\theta\theta'}\,
[(0,-1)(1,0)(0,1)]
\end{align*}

\begin{table}
\centering
\begin{tabular}{|c|c|}
\hline
Representation  & Multiplicity \\
\hline $ \Yasymm_a$
 & ${1\over 2}\left(\pi_a\circ \pi'_a+\pi_a \circ  \pi_{{\rm O}6} \right)$  \\
$\Ysymm_a$
     & ${1\over 2}\left(\pi_a\circ \pi'_a-\pi_a \circ  \pi_{{\rm O}6} \right)$   \\
$( a,{\overline b})$
 & $\pi_a\circ \pi_{b}$   \\
 $(a,b)$
 & $\pi_a\circ \pi'_{b}$
\\
\hline
\end{tabular}
\vspace{2mm} \caption{Chiral spectrum for intersecting D6-branes.}
\label{table}
\end{table}
The chiral matter spectrum is given by the
intersection numbers of table (\ref{table}). Through out the paper we associate to  positive
intersection number $\pi_a \circ \pi_b$ to matter
transforming as the bifundamentals of $(a,{\overline b})$. See also comments in the introduction.
 Given two branes $a$ and $b$ intersection numbers for the bulk part
are defined as
\begin{align}
 \pi^B_a \circ\pi^B_b = 4 \prod^3_{i=1} (n^i_a
\widetilde{m}^i_b-n^i_b \widetilde{m}^i_a)\label{intersection
formula bulk}
\end{align}
and for the twisted sector they are defined as
\begin{align}
 \Big[\alpha^{g}_{ij}\times (n^{I_g}_a,\widetilde{m}^{I_g}_a) \Big]\circ \Big[\alpha^{h}_{kl} \times (n^{I_h}_b,\widetilde{m}^{I_h}_b)\Big] =
 4 \,\delta_{ik} \,\delta_{jl}\, \delta^{gh} \,(n^{I_g}_a \, \widetilde{m}^{I_h}_b - n^{I_h}_b \,
 \widetilde{m}^{I_g}_a)\,\,.
 \label{intersection formula twisted}
\end{align}
Furthermore, N=1 supersymmetry is preserved by the branes as long as
each brane satisfies
\begin{align}
{\widetilde m}^1\,{\widetilde m}^2\,{\widetilde m}^3-\sum_{I\neq
J\neq K}\frac{n^I\,n^J\, {\widetilde m}^K}{U^{I}\,U^{J}}=0
\label{susy1}
\end{align}
and
\begin{align}
n^1\,n^2\,n^3-\sum_{I\neq J\neq K}{\widetilde m}^I\,{\widetilde
m}^J\, n^K\,U^{I}\,U^{J}>0\,\,, \label{susy2}
\end{align}
where $U^I$ denotes the complex structure modulus $U^I=R^I_Y/R^I_X$
of the $I-th$ torus with radii $R^I_X$ and $R^I_Y$.



\newpage

\end{document}